# An Acoustic Emission Activity Detection Method based on Short-Term Waveform Features: Application to Metallic Components under Uniaxial Tensile Test

Fernando Piñal-Moctezuma[a*], Miguel Delgado-Prieto[a] and Luis Romeral-Martínez[a]

[a]*Grup de Motion Control and Industrial Applications (MCIA), Departament d'Electrònica, E.S.E.I.A.A.T. Universitat Politècnica de Catalunya, Colom 11, E- 08222 Terrassa, Spain.*

**Abstract**

The Acoustic Emission (AE) phenomenon has been used as a powerful tool with the purpose to either detect, locate or assess damage for a wide range of applications. Derived from its monitoring, one major current challenge on the analysis of the acquired signal is the proper identification and separation of each AE event. Current advanced methods for detecting events are primarily focused on identifying with high accuracy the beginning of the AE wave; however, the detection of the conclusion has been disregarded in the literature. For an automatic continuous detection of events within a data stream, this lack of accuracy for the conclusion of the events generates errors in two critical aspects. In one hand, it deteriorates the accuracy of the measurement of the events duration, truncating the span of the event (undesirable in evaluation applications), and in the other hand, it causes false detections. In this work, an accurate and computationally efficient AE activity detector is presented, using a framework inspired by the area of speech processing, and which provides the required indicators to accurately detect the onset and the end of an AE event. This is achieved by means of a threshold approach that instead of directly operates with the transduced voltage signal it does so over the Short-Term Energy and the Short-Term Zero-Crossing Rate measures of the signal. The STE-ZCR method is developed for an application related to the continuous monitoring of a single AE channel derived from the characterization of metallic components by means of an uniaxial tensile test. Additionally, two experimental test-benches are implemented with the aim to quantify the accuracy and the quality of event detection of the presented method. Finally, the obtained results are compared with four different techniques, representing the current state of the art related to AE activity detection.

## 1. Introduction

Due to high demanding specifications on safety and reliability demanded by the transportation sector, the design and manufacturing of metallic components and particularly for this end, the accurate mechanical characterization of these materials, represents a critical aspect in which a great deal of technical and scientific efforts have become essential [1,2]. In this regard, one of the most used methods is the tensile test [3,4], in which some important mechanical properties as the Young's modulus, the Poisson's ratio and the yield strength point are measured. One of the outcomes of the assay is the strain experienced by the specimen, typically measured by means of strain gauges and extensometers. Nowadays, in order to complement the accuracy delivered by these devices, some efforts has been made in order to implement additional instrumentation to the test. Such is the case of video extensometer systems, allowing a forensic analysis based on image processing; nevertheless this approach presents two main restrictions, first, the frame digitization period which usually is in the range of unit of milliseconds, and second, the limitation to the surface monitoring which implies a significant loss of information about the internal material dislocations [5]. An alternative mechanical descriptor considered to be added to the test [6], is the analysis of the Acoustic Emission (AE) phenomenon. Which is the manifestation of transient elastic waves in a material, produced by irreversible changes in its crystalline structure. This has been exploited as a potent mean to assist in the detection, location or evaluation of damage.

For the AE as an assessing damage tool, the processing chain is usually composed by the transduction and acquisition of the phenomenon and in the separation and analysis of each captured AE wave. Particularly, for the separation stage of AE events, due to the inherent features of the phenomenon, the resulting waveform from the acquired signal implies a very challenging task in which to be able to perform a proper identification and separation of each AE event; this mainly caused by exhibiting a highly varying background noise, a large difference of amplitudes between events, and a randomness on the incidence and lifespan of these AE events.

Traditionally, the detection work is carried out by the comparison of the acquired electrical signal against a predefined voltage threshold level, in which whenever the electrical signal rises above of this fixed level, it is said that an AE event has been detected. This method, implemented in the early days of using AE as a damage detection tool, emerged due to the then-contemporary lack of available digital hardware capable to process the payload from the large data stream required for a proper digital treatment of the practically baseband signal [7]. Nowadays, after the advent of digital platforms and given its relative efficiency and ease of implementation, nearly all the established standards for AE [8,9], as well as the commercially available instrumentation (and in consequence the majority of the fieldwork), have inherited this voltage thresholding design paradigm as the default method for detecting AE activity. Nevertheless, this method does not directly deal with the abovementioned particularities of the acquired waveform, yielding to inaccuracies on the onset and endpoint determinations related to the AE events; and as in consequence, eventually causing a degradation of the information obtained from the assessing process.

Efforts have been made in order to overcome the aforementioned limitations of the classical thresholding approach, where methodologically, most of the developed methods carry out a transformation of the raw electrical signal into a Characteristic Function (CF) with the aim of emphasizing the presence of AE events, thus enabling a more efficient detection work. Based on non-parametric signal processing methods, the Time-Frequency distribution analysis represents a more accurate tool for detecting the temporal onset of the AE waves; currently most of the efforts are focused on the use of the Wavelet Transform (WT) [10–14], due to improving the resolution of the energy localization of the AE event in the Time-Frequency plane and in consequence having a better accuracy for the onset determination of the wave. Most of these advanced techniques find their inspiration into the Geophysics discipline due to the similar production between phenomena, such is the case of the Short-Term Average to Long-Term Average ratio (STA/LTA) [15], developed for determining earthquake events while maintaining a low count of false-positive alarms. The most revised technique in the AE area, is the Akaike Information Criterion (AIC) [16], that models the time series data at the beginning of the AE raw signal under an autoregressive scheme (of low order), with the aim of achieving an estimation of two locally stationary parametric components of the framed original signal (noise and AE activity), hence allowing the identification of the AE onset.

Despite of these advanced methods clearly represent a superior alternative to the classic thresholding technique, traditionally they have evolved in light of applications for locating AE sources, where a highly accurate onset detection is critical, so consequently, the issues related to the endpoint determination have been disregarded. Methodologically, this has implied that instead of considering, some intrinsic feature related the phenomenon, all the AE activity detection methods only make use of the combination of a threshold level along with a fixed timer in order to determine the conclusion of the AE event. Nevertheless, due to the stochastic manifestation of said events, this methodology leads to inaccuracies on the measurement of the endpoint determination, what directly affects to the quality of detection of all methods [17] (i.e., the amount of properly detected AE events on a survey); being a critical aspect for assessing damage applications.

A last matter to considering is that owing to the high data rates required to process the AE phenomenon, these advanced methods are computationally expensive, so usually they are implemented under an offline framework (first capturing the data from the survey, to later separating the AE events). Although some efforts has been made in order to implement hardware architectures that can operate under an online approach [18–20], strategies that can lead to faster and efficient implementations (this particularly necessary for long surveys), are desirable.

Evidently, Acoustic Emission is not the only discipline related to the Signal Detection Theory (SDT). Particularly in the speech processing discipline, SDT finds its application on the Voice Activity Detection (VAD) stage [21,22]; where a randomly present speech activity from a highly noisy digitized signal aims to be extracted in order to reduce the payload from the subsequent stages given a particular application (e.g., voice telecommunications, artificial intelligence, hearing aids among others). One of the best-established automatic VAD for the speech processing area is the technique developed by Rabiner and Sambur [23]. This parameter-based VAD was originally designed with the objective to accurately detecting the beginning and the end of an utterance, while preserving an efficient and straightforward processing scheme as well as being robust against varying background noise. This was achieved with the use of two indicators of the signal: the Zero-Crossing Rate (i.e., the rate at which the waveform changes from a positive to negative voltage and back), and the Short-Term Energy. Additionally, the algorithm was intrinsically capable of executing suitably in any realistic acoustic environment in which the signal-to-noise ratio (SNR) was in the order of 30dB. Despite of the dissimilar origins between Acoustic Emission and Speech phenomenon, and as in consequence the technical requirements in order to be processed (e.g., instrumentation, bandwidth —8kHz vs. 2MHz, etc.,), the behaviors of their waveforms share similar characteristics, such is the case of a high variance on the occurrence of activity, rapid varying background noise and significant dynamic range.

In this work, an AE activity detector inspired by the VAD developed by Rabiner and Sambur [23] is presented. The detector is revised for an application related to the recording of a single channel from a continuous AE monitoring, derived of the characterization of a metallic component by means of an axial tensile test, where the AE waves derived from this assay typically exhibit a large difference of amplitudes between events, a stochastic occurrence and duration, and a highly varying mechanical background noise due to cumulative reflections.

In order to evaluate the performance of the presented STE-ZCR method, two experimental setups are arranged: a) a collection of Hsu-Nielsen sources with aim to quantify the accuracy of the onset and endpoint determinations as well as for assessing the robustness of the method with regard to induced background noise. b) The continuous detection of AE events from an AE data stream obtained from a standardized tensile test in order to quantify the quality of detection of this method. Additionally, with the aim to establishing a common benchmark of comparison with some of the advanced detection techniques currently available in the literature, the results of the presented method are compared with: a) A classical threshold technique enhanced by means of the Instantaneous Amplitude [24], b) A typical STA/LTA detector [25], c) A two-step AIC picker [26], d) An a CWT-Otsu detector over a binary image mapping [14] which alike c) uses a modified Allen's Formula as CF for the threshold-based early detection.

The contribution of this work is to provide an automatic AE event detector for a continuous data stream, which excels the detection capabilities (i.e., the onset and endpoint measures, as well as the detection quality of events), with regard to the current methods present in the literature. Novelties of this work include the use of the Short-time analysis, which offers an accurate, computationally efficient, and noise resilient framework to detect AE events. In addition, the use of an indicator obtained from the waveform of the signal with the aim to determine the conclusion of an AE event, instead of the traditional combination of a threshold level and a fixed timer.

This paper is organized as follows: In **Section 2** the proposed STE-ZCR detection method is introduced. **Section 3** describes the two experimental setups with the aim of benchmarking the method and comparing its performance with the current advanced methods from literature. In **Section 4,** results are presented and discussed. Finally, in **Section 5** conclusions are provided.

## 2. Methodology: AE Activity Detector

The idea behind of a Short-Time or Short-Term Analysis (ST-ANLYS) relies on the stationarity of a time series, with aim of creating a new sequence that can represent some varying feature of the original. This is achieved starting from the fact that some signals (such is the case of an AE signal) intrinsically will not show a stationary behavior, that is, during their lifespans there will not be a clear tendency of repeatability on them (e.g., statistical mean and covariance). However, some other signals (again, an AE signal) when are enough and equally time segmented will show a relative slow variation (compared with the original time frame) for some property between segments, so these time segments (usually known as analysis frames) relate the analysis of the signal regarding a fixed size time window.

An activity detector exploits this artificially induced stationarity by identifying the relatively higher energy and the less number of zero crossings that are associated with a performing phenomenon, contrary to an idle activity condition. Despite of the inherent uncertainties induced by a ST-ANLYS, it has proven to be an efficient tool with the aim of identifying the regions of activity of a signal.

For this work, the STE-ZCR method (see **Fig. 1**), is composed at first of the generation of two characteristic functions by means of the ST-ANLYS of the Energy and the Zero Crossing Rate respectively of the acquired AE signal; and second, taking as inputs these pair of CF's, the detection of AE events performed by a dedicated algorithm.

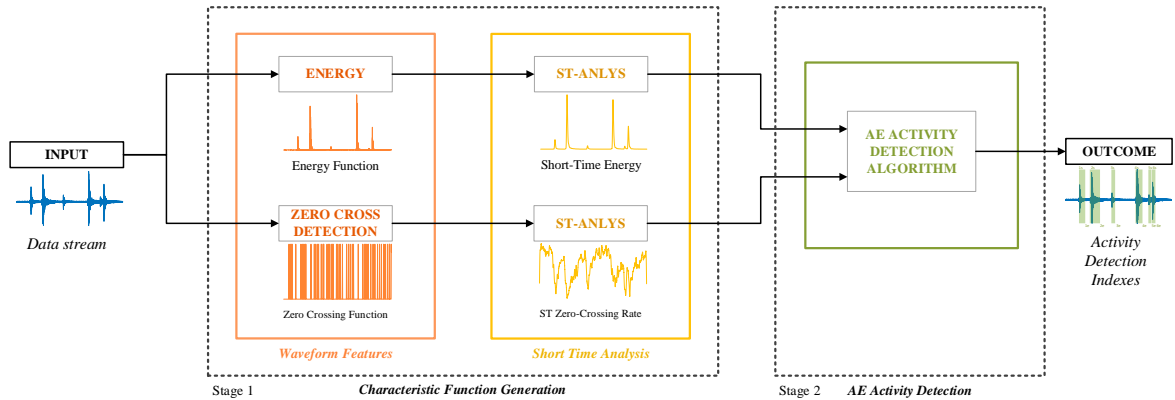

**Fig. 1.** Block diagram for the STE-ZCR method, composed by two main stages: 1) Production of two CF by means of the ST-ANLYS framework, and an 2) Activity detection algorithm that searches for the onset and endpoint by means of the STE and the STZCR respectively. The outcome of this method is a set of pair indexes that mark the temporal start and end sampling points with regard to the data stream at the input, for each detected AE event.

*2.1 Stage 1. Characteristic Functions Generation: Short Time Analysis Framework*

*2.1.1 Short Time Energy*

As is known, the energy $E$ of a discrete-time signal of length $L$ from a point of view of signal processing can be expressed as $\sum_{m=0}^{L}|x(m)|^2$. For this framework, the short-time energy of the signal is defined as:

$$E_{\hat{n}} = \sum_{m=\hat{n}-N+1}^{\hat{n}} |x(m)|^2 \, w(\hat{n}-m) \tag{1}$$

where $w(\hat{n}-m)$ is a window function of $N$ width, centered at sample $\hat{n}$, and where $\hat{n}$ in turn indicates the overlapping factor between windows through the relation $\hat{n} = kT$, with $k = 0, 1, ...,$ and $T < N \leq L$.

For this AE activity detection application, the association of $E_{\hat{n}}$ lies in provide a measure to separating the presence of AE waves from idle activity on the acquired sequence, since values of $E_{\hat{n}}$ for an AE wave are considerably greater than noise floor energy. It is important to note that under this scheme there are three different parameters to configure for the short time energy approach:

a)   *Window function.* As is well known, this type of functions typically must meet some requirements in order to be considered suitable for the processing of the signal (i.e., smoothness, non-negative terms, compact support, square integrable resultant products). The resulting waveform from the analysis will strongly depend on the choice of the window function, and since there is not exists an optimal overall option, selection must entirely rely on the scope of the application in which the analysis will perform (e.g., spectral, statistics, etc.,) and therefore on the pursued feature to highlight. For applications related with analysis of transients (like an activity detector), where the objective is to accurately concentrate the energy of the signal in the time domain taking advantage of the low-pass filtering nature of the ST-ANLYS at expense of diminishing the bandwidth of the resulting signal, typical windows include rectangular and raised cosine categories.

b)   *Window length.* Ideal response of any CF is to depict a feature of interest at a rate comparable of the original signal; the resulting signal by means of a ST-ANLYS will inherently contain uncertainties on the temporal relocations of the analyzed feature, still, these can be significantly reduced through a proper choice on the length of the window. As in the case of the window function, the length of the window relies entirely on the application, there is not ideal generic value, thus it is necessary to take considerations about of the tradeoffs on the responsiveness in the selection of the window length with regard of the expected lifespan of an AE wave in function of the material under analysis:
   i. <u>Small length, uncertainties due to a small amount of data</u>. Although is desirable to have a CF function that rapidly responds to abrupt changes of the signal under analysis, a too short window will not reveal any stationarity derived of the ST-ANLYS.
   ii. <u>Medium length, uncertainties due to loss of rapid transients</u>. While a conservative value for the analysis will provide a superior depiction about of the stationarity of the signal (aside from discarding fast mechanical noises), this will not accurately resolve the rapid transitions between a pair (or more) of too near AE waves (also known as cascaded hits).
   iii. <u>Extended length, uncertainties due to significant amount of exclusions of signal changes.</u> Even if what is sought as result of the ST-ANLYS is to obtaining a smooth CF capable to easily portray the tendency of the analyzed signal (this achieved through increasing the length of the window size), an excessively wide window will suppress the dynamics of the signal making difficult to identify sharp changes.

c)   *Window overlapping factor.* Once that the type and the length of the window have been selected, the last step of the ST-ANLYS is to slide the window over the analyzed signal in order to reveal the stationarity of the signal. For applications where the extracted ST-ANLYS between subsequent samples may not be required since the variation of the feature is relatively slow, the shift can be kept larger than one sample. Under this overlapped scheme, the computational load of the analysis can be largely reduced (this is particularly useful when long duration signals are under evaluation); typical overlapping values are in the range of 50 to 75% of the window length. Still, since overlapping implies to downsampling the resulting signal by a $T$ factor (see **Eq. 1**), for applications where a high temporal accuracy on redistributing a feature is required, is desirable to keep the sample shift as minimum as possible.

### 2.1.2 Short Time Zero Crossing Rate

The zero-crossing point for an alternating electrical signal is the instantaneous time value when voltage equals to zero. Since the point of view of the discrete-time signal processing, a zero-crossing point occurs where two adjacent sampling points on the sequence have different mathematical signs (i.e., having opposite polarities). The ratio count of zero crossings over a unit of time is a quite basic but still effective measure of approximating the spectral content of the signal. The short-time zero-crossings rate per sample in this work is defined as:

$$Z_{\hat{n}} = \frac{1}{2N} \sum_{m=\hat{n}-N+1}^{\hat{n}} |sgn(x[m]) - sgn(x[m-1])| w(\hat{n}-m) \tag{2}$$

where $w(\hat{n} - m)$ is the chosen window function, $N$ is the length of said window centered at sample $\hat{n}$, and also indicating the overlapping factor between windows through the relation $\hat{n} = kT$, with $k = 0, 1, ...,$ and $T < N \leq L$. Finally $sgn$ denotes the signum operator defined as:

$$sgn(x[n]) = \begin{cases} 1, & x[n] \geq 0 \\ -1, & x[n] < 0 \end{cases} \tag{3}$$

Still, is possible to express the ZCR normalized for an interval of $M$-samples, whence ZCR becomes:

$$Z_M = M Z_{\hat{n}} \tag{4}$$

and where an interval of $\tau$ seconds corresponding to $M$-samples is:

$$M = \tau F_s \tag{5}$$

As in the case of the STE, in order to obtain a proper description of the STZCR it is necessary to make the same considerations about of the type, length and the overlap shifting of the window with regard to the required application. Additionally, there are further practical considerations to make (by means of a filtering scheme) before applying the analysis, since ZCR is heavily biased by DC offset of the analog-to-digital conversion, the 50/60 Hz mains hum and low frequency mechanical background noises.

### 2.2 Stage 2. AE Detection Algorithm

Once that both CF's (STE and ZCR) are obtained, the second stage of the STE-ZCR method is to find the pairs onset/endpoint for the AE events on the sequence. For this, using the fact that the waveform derived from an AE event will exhibit higher energy and lesser ZCR count, it is possible to set up the basis to implement an algorithm that can detect AE events on a straightforward but yet efficient scheme (see **Fig. 2**).

The algorithm makes use of two different fixed threshold values for its operation, the *Identification Upper Threshold* (ITU) that works on the STE signal in order to detect new AE events and the *Identification Zero Crossing Threshold* (IZCT) that works on the STZCR with the aim to determining the endpoint for a detected AE event. Thus, a previous characterization of the instrumentation with regard to the surveyed material is highly recommended to characterize the background noise level and the amplitude of the electrical waveform of the monitored AE channel, in order to carry out a proper calibration of said threshold parameters. In addition to the CF's generated by means of the ST-ANLYS, the algorithm makes use of the derivative of the STE with the aim to refine the onset determination of the AE event. Finally, with the purpose to enhance the quality detection, a basic adaptive threshold scheme is implemented by continuously taking measurements of background noise and adjusting the threshold levels between search jobs iterations.

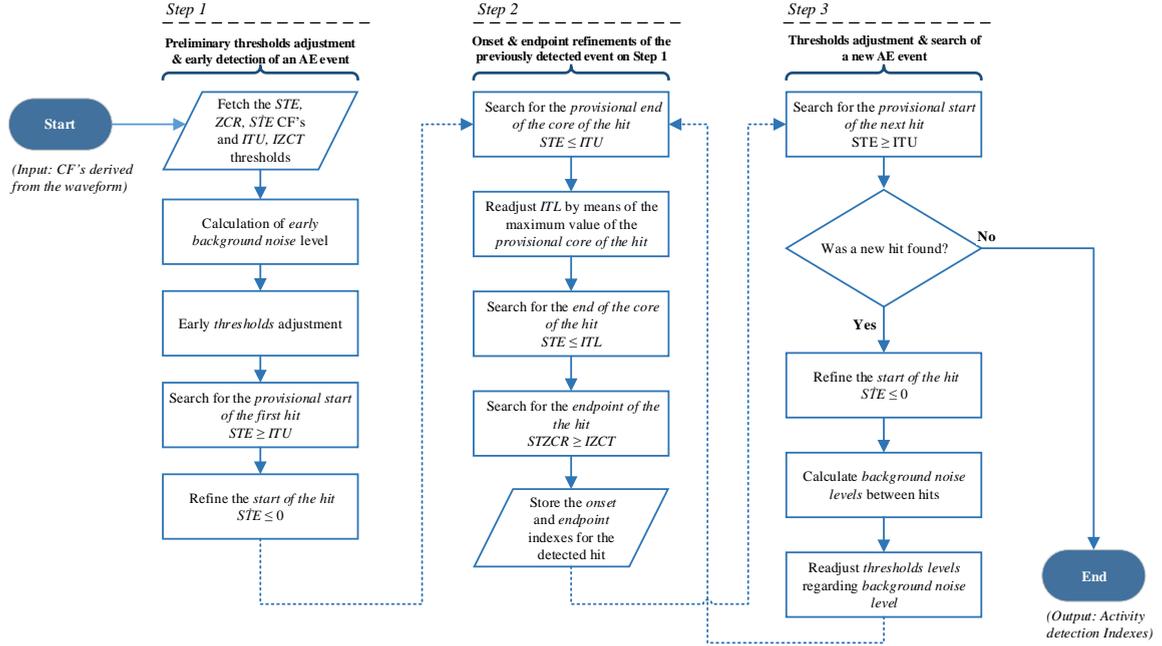

**Fig. 2.** Flowchart for the activity detection algorithm composed of three main steps: 1) Preliminary calculation of background noise for an idle state of the signal, 2) Search job for the onset and endpoint of a hit, 3) Threshold level adaptation regarding to the updated background noise. Reader is also referred to Fig. 3 for a visual instance of the CF's and threshold levels used by the algorithm.

The search work initiates with an estimation of the *STE early background noise level* for a small segment of the sequence, and belonging to the beginning of said sequence (where is assumed that there is not exists any AE hit yet). This is achieved by means of the sum of the arithmetic mean $\bar{x}_{STE}$ and the α-weighted factor for the standard deviation $\sigma_{STE}$ of the STE signal, (this α factor can be estimated along with the previous calibration for the thresholds, if a heavy background noise is expected the value for this weighting value must be incremented). In order to perform the first threshold adaptation, the *early background noise level* is added to the preset *Identification Upper Threshold* (ITU), expressed in levels of energy ($v^2 \cdot s$) $ITU_{adjust} = ITU + (\bar{x}_{STE} + \alpha * \sigma_{STE})$.

For the calculation of the *STZCR early background noise level*, the same calculations that for the STE case are performed, now over the same segment length belonging to the STZCR sequence. Thus, the resulting *adjusted Identification Zero Crossing Threshold* will be: $IZCT_{adjust} = IZCT + (\bar{x}_{STZCR} + \alpha * \sigma_{STZCR})$, expressed in normalized ST-ZCR of the signal, for a length window of $M$ samples ($\widetilde{STZCR}$). Additionally for this parameter, for a simpler threshold presetting, it is also possible to handle the *IZCT* threshold value as a percentage of the computed *early background noise level*.

After both parameters ($IZCT_{adjust}$ and $ITU_{adjust}$) are computed, the next step is to find the first sample $n_{Prov-onset}$ in the *STE* sequence where: $STE[n] \geq ITU_{adjust}$. This $STE[n_{onset}]$ sample will be the *provisional onset hit*.

The last task for this first stage comprises the refinement of the onset detection of the hit by means of the first derivative of the *STE* sequence ($S\dot{T}E$). The purpose on the use of this signal is to take advantage of the sensitivity to the energetic change that the STE provides, by finding the first incidence of the energetic variation of the captured AE phenomenon. Thus, it is possible to assume that in this sample the arrival of p-waves is manifested. For this is carried out a backward search from the corresponding $S\dot{T}E$ sample of the *provisional onset hit* sample until finding the $n_{true-Onset}$ sample where $S\dot{T}E[n_{true-onset}] \leq 0$. This sample is indexed as the *true onset hit detection*, and will be the first outcome of the onset-endpoint pair of indexes.

The first task for the second stage of the algorithm consists in finding the *core of the hit*. Since in this region most of the AE wave energy is concentrated, it is necessary to delimit it accurately with the aim to aid to find the lifespan of the hit. For this step, the first subtask requires to find the *provisional end of the core of the hit* $n_{provEoC}$, this can be picked up readily by locating the first sample from the *provisional onset hit* sample where $STE[n] < ITU$.

Next, for the search work of the *end of the core of the hit*, it is required to find the maximum value $STE_{max-core}$ regarding to the provisional core of the hit (i.e., the range between *STE*[*provisional onset hit*] and *STE*[*provisional end of the core of the hit*]), to then readjust the *Identification Lower Threshold* (*ITL*) expressed in levels of energy ($v^2.s$). For this, it is necessary to start from the fact that for the Acoustic Emission discipline one of the most accepted models [27–34] of an electrically transduced AE wave, considers to the wave as an underdamped sinusoidal function with the form:

$$u(t) = \begin{cases} A \exp\left(\frac{-t-T}{\gamma}\right) \sin 2\pi v_0(t-T), & t \geq T \\ 0, & t < T \end{cases} \quad (6)$$

Where $A$ is the amplitude and $T$ is the arrival time of the AE wave; and $\gamma$ and $v_0$ the decay constant and the resonant frequency both belonging to the sensor. Therefore, the envelope of the wave of **Eq. 6** can be expressed as:

$$e(t) = Ke^{-v_0 t}, \quad t \geq T \quad (7)$$

Is evident that this simple exponential decay model can be enhanced, still, for the application of this work it is only required to identify that the STE CF corresponding to an AE wave behaves as an impulse response function of a linear-time-invariant dynamical system of first order. As is well known, the time constant $\tau$ that characterizes the response of the system and its bandwidth for a system like in **Eq. 7**, it is located at the instant of time $t_\tau$ where $e(t_\tau)$ equals to 36.8% of its maximum value. Therefore, taking advantage of this fact, it is possible to assume that the core of an AE wave will expire when the STE equals to the 36.8% of its maximum value. Hence, the value for the *ITL* threshold will be adjusted to $ITL = 0.368 STE_{max-core}$ to then perform a forward search from $STE[n_{max-core}]$ until find the $n_\tau$ sample where $STE[n] \leq ITL$. This $n_\tau$ sample will be the *end of the core of the hit*. While it is true that after the *end of the core of the hit* most of the energy of the AE wave is nearly vanished, the region between of this *STE* endpoint detection and the one that will be determined through the *STZCR* CF still will comprise a relevant content of energy. Thus, the joint use of these pair of CF's will provide a more robust and precise approach for determining the end of the lifespan of an AE wave. For this, a simple search forward work over the *STZCR* sequence will be performed from the $n_\tau$ sample (related with the *end of the core of the hit*) until finding the $n_{endpoint}$ sample where $STZCR[n] \geq IZCT_{adjust}$. This sample will be indexed as the *true endpoint of the hit*.

With these two indexes as outcome (*true onset hit detection* and *true endpoint of the hit*) concludes the work for the second stage of the algorithm and the third and last part of this will be executed. Searching for this a new hit over the *STE* sequence, and using the last adjusted value of the *ITU* threshold from the *true endpoint of the hit* sample. Once detected, a refinement of the onset detection will be performed.

Next, an update of the noise levels over their corresponding CF's will be performed (from the last *true endpoint of the hit* to the new *true onset hit detection*), to then readjusting of the *ITU* and *IZCT* thresholds values. However, with the purpose to avoid an overlap of AE events, if a new hit were detected before the *true endpoint of the hit* sample (obtained by means of the STZCR CF), the *true endpoint of the hit* must be readjusted to one sample before of the *true onset hit detection*. Finally, the second stage of the algorithm will be performed again and the iteration repeated until all events of the sequence under analysis are detected.

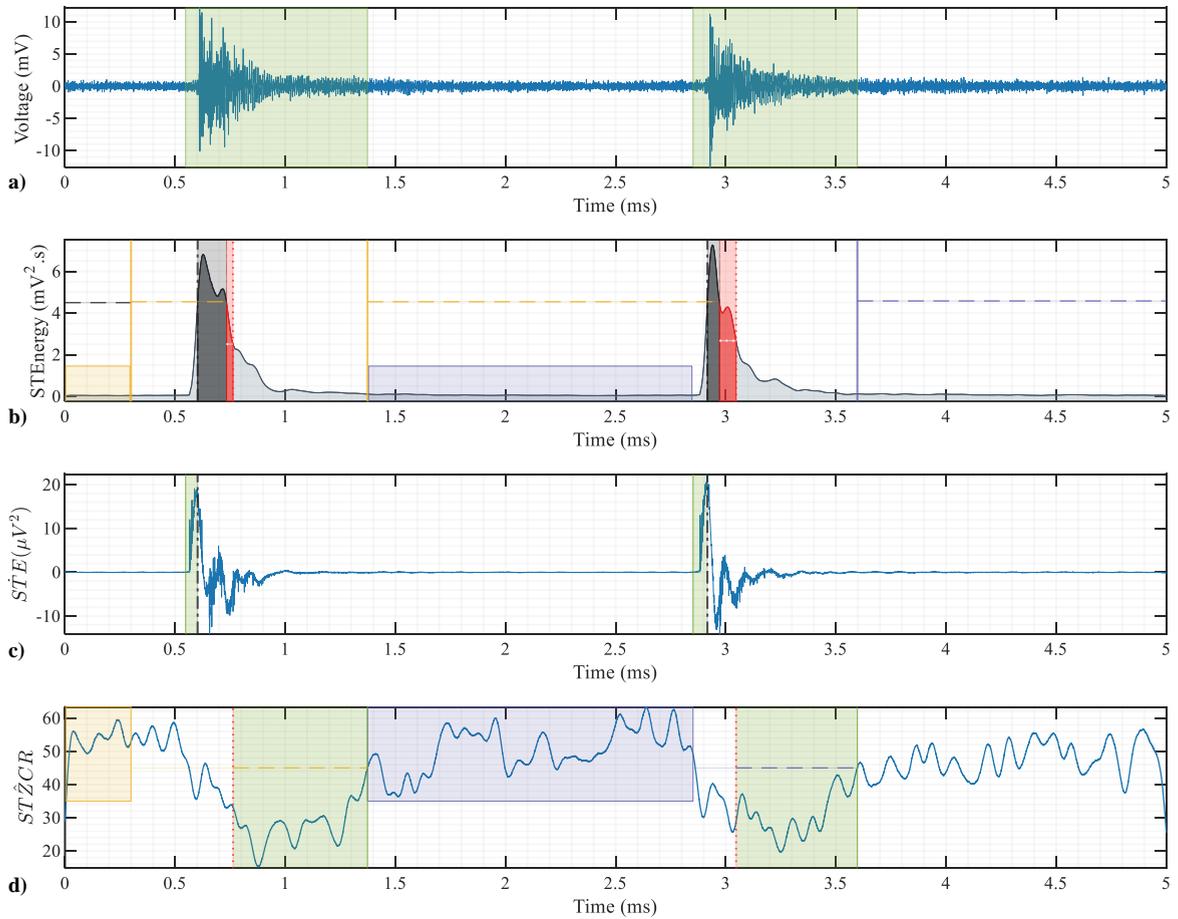

**Fig. 3.** Depiction of the STE-ZCR detection method for a data 5ms frame, which contains two AE events. **(a)** Two AE events and their corresponding lifespans detected by the AE activity detector (green shaded areas). **(b)** Onset detection work. Short-time energy CF (blue-steel area under the curve), preset threshold level (horizontal dotted black line), signal segment for the early background noise calculation (yellow shaded area), first adjusted threshold (horizontal dotted yellow line) and sample of activation (vertical solid yellow line), signal segment for the second background noise calculation (lilac shaded area), second adjusted threshold (horizontal dotted lilac line) and sample of activation (vertical solid lilac line). Provisional core of the hit (gray shaded areas), Identification Lower Threshold (horizontal dotted white line), signal segment for the end of the core search job (red shaded areas). The core of the hit for each AE event is composed by the gray and red shaded areas respectively. **(c)** Refinement work for the onset detection. Backward search job from the start of the core sample (vertical dash-dotted black line) through the derivative of the STE-CF (blue solid curve). Onset time samples (vertical solid green lines). **(d)** Endpoint determination work. Short-time ZCR CF (blue solid curve), signal segment for the early background noise calculation (yellow shaded area), samples of activation *end of the core of the hit* (vertical dotted red lines), first adjusted threshold (horizontal dotted yellow line), signal segment for the second background noise calculation (lilac shaded area), second adjusted threshold (horizontal dotted lilac line), endpoint determination samples (vertical solid green lines).

## 3. Experimental procedures

Following the experimental procedures presented in the literature, two experimental scenarios were arranged in order to quantifying the competency of the proposed STE-ZCR method in front of a comparative common framework with regard to some of the most significant methods for detection of AE events from literature.

First experimental scenario is prepared to quantifying the precision of the measures regarding to the onset, endpoint and lifespan determinations. For this, a collection of one-hundred different AE waves derived from a standardized Hsu-Nielsen test is processed and their corresponding detection absolute errors are calculated. Then, with the purpose of evaluating the operating robustness in front of noise, each AE synthetic wave of the dataset is tainted with three different levels of Additive white Gaussian noise (AWGN), and their corresponding detection absolute errors are newly calculated. Additionally, in order to evaluate the throughput of the presented framework, the processing time of each analyzed AE event is measured.

The second test-bench involves to measuring the quality of detection by means of the accuracy, precision, sensitivity, f1-score, false-discovery rate and false-negative rate statistical indicators. For this, a data frame derived from a tensile test of a metallic component containing an ample variety (in duration, amplitude and incidence) of AE waves is processed, and the amount of properly detected AE events is totaled. The required processing time with regard to the data frame is also calculated, as well as the absolute errors of the onset, endpoint and lifespan detections of the corresponding true-positive events.

For both experimental scenarios, one sensor (Physical Acoustics WSα, 100-1000 kHz) was attached to the surface of each corresponding metallic component (using a silicon-based couplant agent). The resulting electrical signals were amplified (by a Mistras preamplifier 2/4/6) with a gain of 20dB (BW 10-2500 kHz). The amplified electrical signals, were recorded under a free-running sampling scheme (using a CSE4444 digitizer of the GaGe manufacturer), with a sampling frequency of 5MHz for the Hsu-Nielsen data and 10MHz for the tensile test data (both samplings with a resolution of 16-bit depth). All the raw signals are band-pass filtered by means of a FIR equiripple implementation (10-2200 kHz). Preliminary to performing the corresponding test-benches, the onset and endpoint times of each AE-wave that will be processed, are manually picked supported by means of time-voltage plots as well as by a time-frequency distribution of high resolution [35].

The test benches as well as the considered methods are implemented using software scripts executed by MATLAB® R2018a in a PC with a CPU Intel Core i7-6800k (3.4GHz) and 64GB of DDR-2400 RAM.

For the state of the art methods used in the experimental scenarios, the most fitting calibration parameters in regard to the test-bench are done following the recommendations of related literature [26,36–40] as well as the current standards [8,9,41–45].

*3.1 Hsu-Nielsen data test-bench*

For the Pencil-lead breakage (PLB) test-bench, and for each of the one-hundred realizations, a graphite lead of ⌀ 0.5mm, 2.5 mm tip-length and with a contact angle to the surface of the specimen of 60° is used. In addition, a distance of 12cm between source and sensor is preserved (see **Fig. 4**).

For a frequency range of up to 1MHz, the average phase velocity for the extensional mode is of 5194m/s and for the group velocity case is about of 4471m/s. In **Fig. 5**, is shown the characterization of the used sheet specimen by means of its dispersion relation of the fundamental Lamb wave modes (obtained using the Wavescope software [46]). With this information, and taking into account the operative frequency of the used sensor, the source-sensor layout shown in **Fig. 4** and assuming an ideal isotropy in the material, it is possible to neglect the effect of the change of velocity for this experiment.

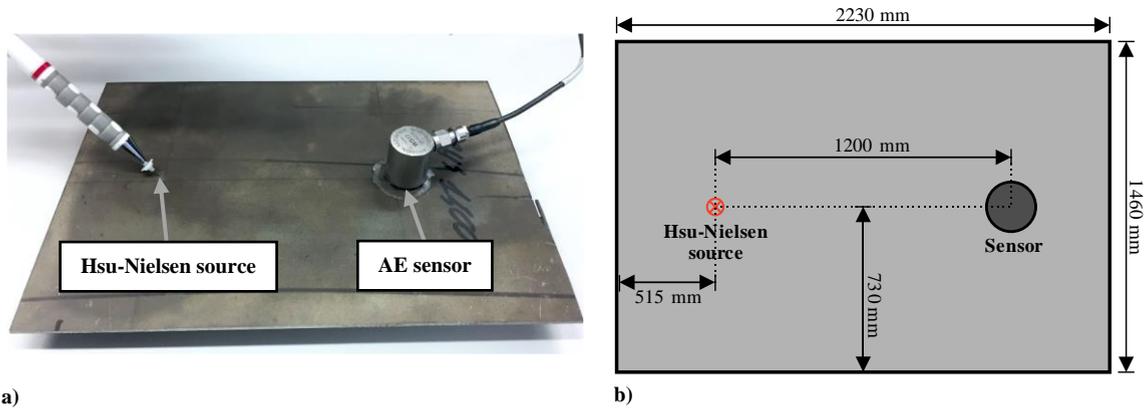

**Fig. 4.** Setup for the standardized Hsu-Nielsen test-bench over a Press-Hardening 1500 steel plate specimen. **(a)** Photograph of the AE sensor, the guide-ring tube used to generate the artificial sources and the steel plate (stood over a foam base). **(b)** Schematic diagram indicating the dimensions of the specimen and the locations of the source and sensor.

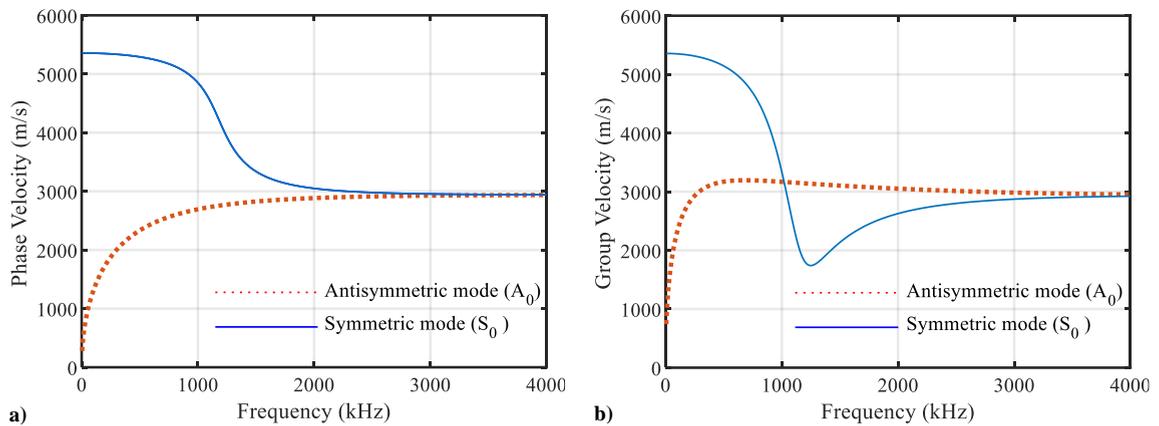

**Fig. 5.** Fundamental dispersion curves of the Press-Hardening 1500 steel sheet with a thickness of 2mm, Young's modulus of 211GPa, density of 7850kg/m$^3$, Poisson's ratio of 0.3 and shear modulus of 83GPa. **(a)** Phase velocities. **(b)** Group velocities.

For repeatability purposes, each synthetic AE wave is edited so its peak value is centered on 5ms and the signal be extended during 40ms more, as a result each AE wave from the data set collection will exhibit an average lifespan of 20.86 ± 1.16ms. A typical waveform obtained from this process is shown on in **Fig. 6**.

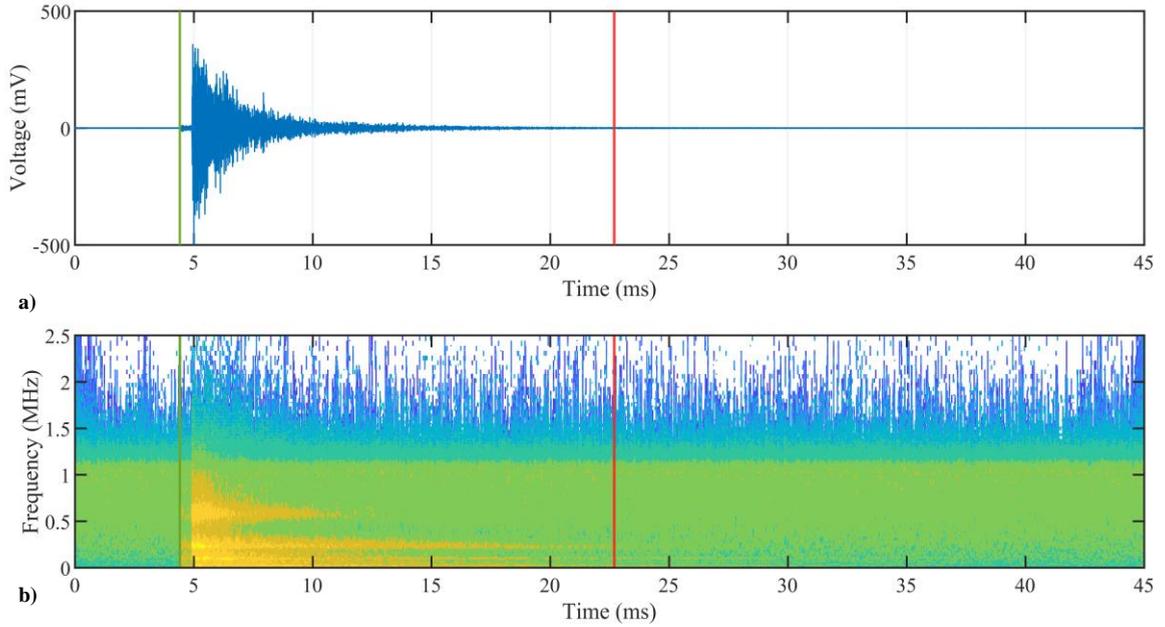

**Fig. 6. (a)** Typical AE waveform event analyzed in the synthetic data test-bench. **(b)** Synchrosqueezed wavelet transform (analytic Morlet wavelet), used to assist in the manual determination of the onset and endpoint locations of the AE wave (green and red lines respectively).

### 3.1.1 Operational robustness in front of background noise

Second objective for this test-bench consists of evaluating the operational robustness of the method in front of background noise. For this, since each AE wave from the data set collection exhibits an average Signal-to-noise ratio (SNR) of 27.1 ± 1.15dB, their corresponding SNR will be decreased by means of AWG noise in three different rounds of analysis of 20, 15 and 10 dB respectively (see **Fig 7**). Then, the relative errors for the onset, endpoint and lifespans are newly calculated for each round, as well as the consumed processing times.

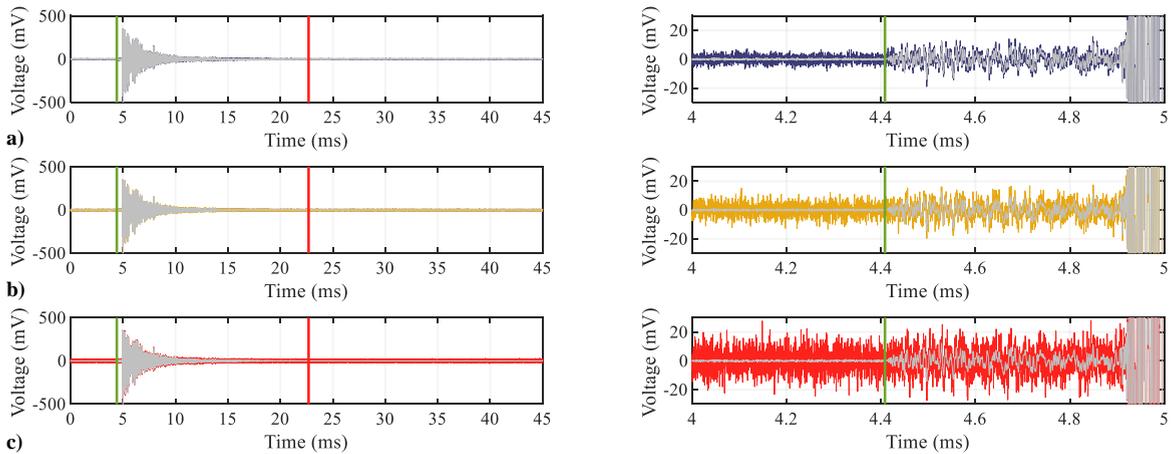

**Fig. 7.** Example of an AE wave used for the evaluation of operational robustness against noise, and onset and endpoint locations (green and red lines respectively). In the test, a synthetic AE signal (gray) is tainted by AWGN in order to obtain three different signals with levels of SNR of: (a) 20dB (lilac), (b) 15dB (yellow), (c) 10dB (red). Images of the right column show a 1ms zoom of the corresponding data frame for the wave onset.

## 3.2 Uniaxial tensile test

The objective for the second test-bench is to quantify the quality of event detection by means of statistical indicators in front of field data. For this, a tensile test of a metallic component is carried out (see **Fig. 8**).

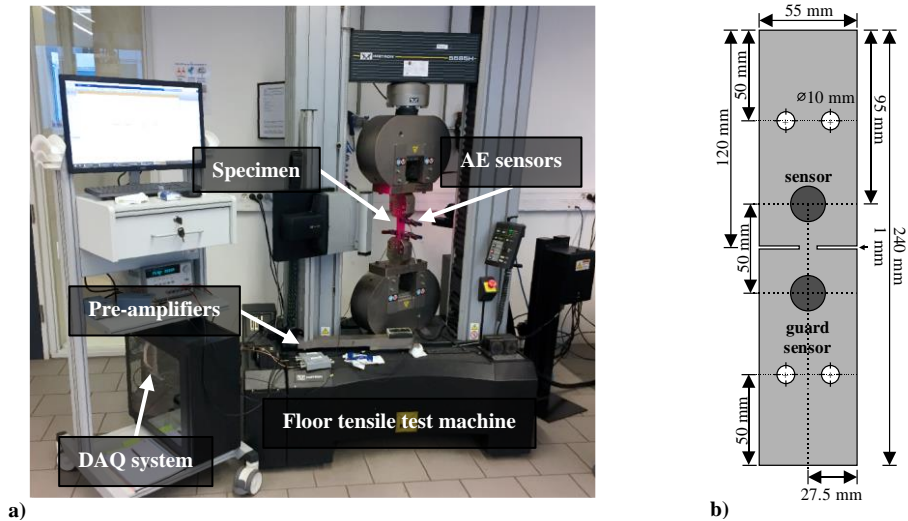

**Fig. 8.** Setup for the tensile test-bench over a Ferrite-Pearlite annealing steel sheet specimen (load rate of test 1mm/min). **(a)** Photograph of the instrumentation and the floor tensile test machine used. **(b)** Schematic diagram indicating the dimensions of the metallic specimen and the locations of the sensors.

Even though a pair of identical sensors were attached during the assay, for this test-bench, only the signal of the main sensor will be analyzed.

As in the case of the Hsu-Nielsen experimental scenario, when is assumed an ideal isotropy in the material, the characteristic average extensional mode wave velocities (~5104m/s for the phase and ~4348m/s for the group **Fig. 9**) and taking into account the operative frequencies range of the sensor (and its location over the specimen, **Fig 8.**), it is possible to neglect the effect of the change of velocity for this experiment.

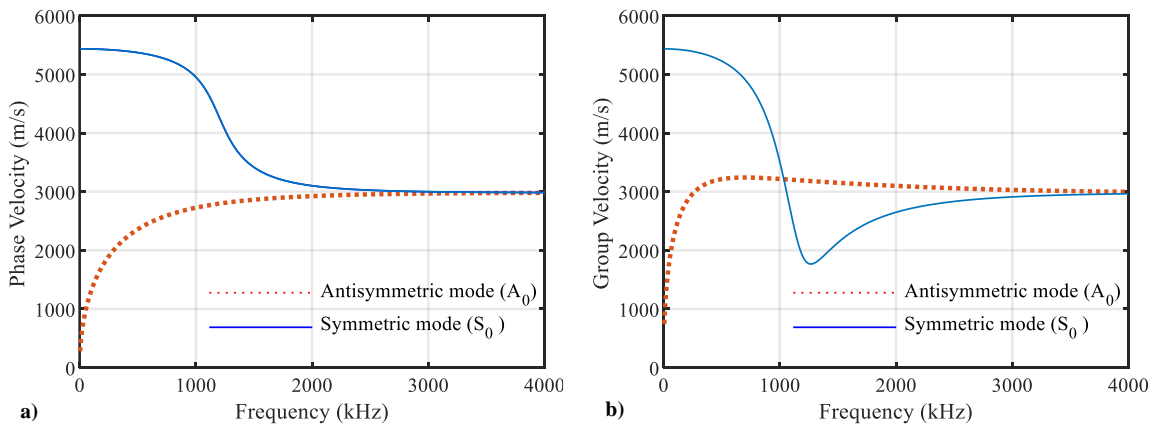

**Fig. 9.** Fundamental dispersion curves of the Ferrite-Pearlite annealing steel sheet with a thickness of 2mm, Young's modulus of 205GPa, density of 7850kg/m$^3$, Poisson's ratio of 0.3 and shear modulus of 83GPa. **a)** Phase velocities. **b)** Group velocities.

The AE signal produced by the tensile test was collected; and for the experimental scenario a frame length of 500ms that contains 380 AE events is used as the input for each detection method (see **Fig. 10**). For each of said AE events, their onset and endpoint locations are manually picked supported by the frame waveform and its corresponding time-frequency distribution.

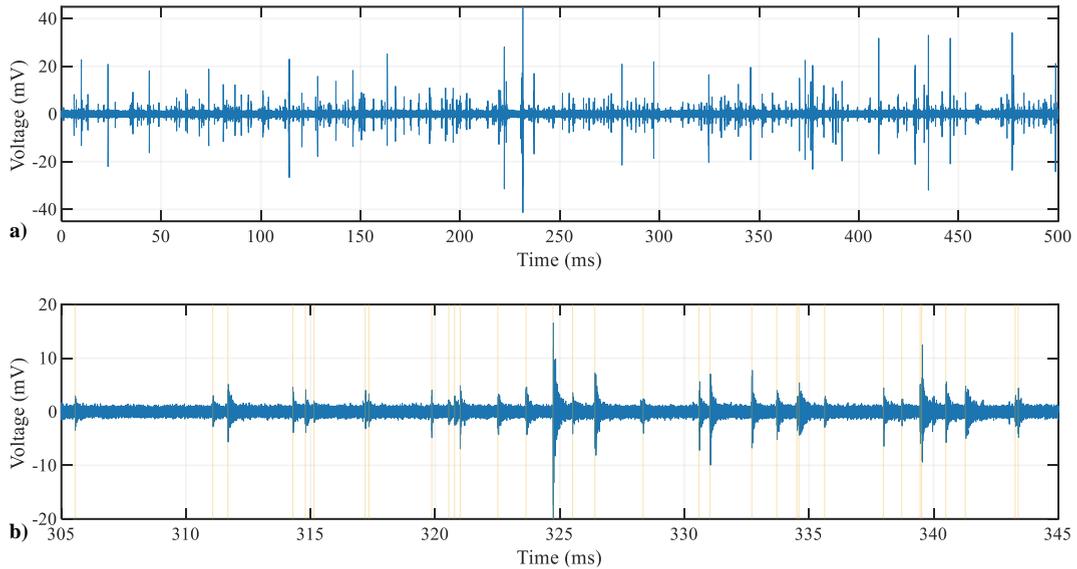

**Fig. 10. (a)** Signal frame used for the field data test-bench. **(b)** Zoom of 40ms of the signal frame, showing the ample variety on the incidence, lifespan and amplitudes of the AE waves present in the test-bench (manual onset picks indicated by the vertical yellow lines).

## 4. Results and discussion

The competency of the STE-ZCR method was analyzed in front of two different experimental scenarios. Additionally, its performance is compared against four different representative AE detection techniques: (a) a classical threshold detector enhanced by means of the Instantaneous Amplitude envelope [24], (b) a STA/LTA detector [25], (c) a two-step Akaike Information Criterion picker [26], (d) and an Otsu detector working over a binary map image based on the Continuous Wavelet Transform [14], (which alike (c) uses the same waveform derived from the Allen's Formula as CF for the threshold-based early coarse detection).

*4.1 Hsu-Nielsen data test-bench. Accuracy of the Onset and Endpoint determinations.*

As aforementioned, the objective for this test-bench is firstly to quantify the accuracy of the measurement for the onset, endpoint and lifespans by means of the absolute error of each measure, and secondly to evaluate the operational robustness of detection in front of background noise.

For the calibration of the STE-ZCR method with regard to the corresponding temporal window analysis (type, length and overlapping factor), after a series of exhaustive trials, it was determined that one of the window functions that accomplished higher accuracy results was the Hamming implementation (in general those belonging to the raised cosine family). For the window overlapping factor, at expense of increasing the computational load, the best accuracy was achieved by maintaining the window overlapping to one sample, (i.e., by directly convolving the instantaneous energy and the window function). Finally, the choice of the duration values of the window time, the threshold levels and the weighting factor for the noise analysis will be entirely determined by a prior calibration of the employed instrumentation as well as by the mechanical properties of the material. The calibration values of the comparative methods (see **Table 1**), was carried out following the recommendations of the related literature [26,36–40] as well as the current standards [8,9,41–45]) of the AE discipline.

**Table 1**
Calibration parameters values used for each method for the Hsu-Nielsen test-bench.

| Parameter | Method | | | | |
|---|---|---|---|---|---|
| | IA | STA LTA | AIC | CWT Otsu | STE ZCR |
| Fixed threshold level | 3e-3 | 5e-4 | 2e-1 | 2e-1 | 2e-4 |
| Hit Definition Time [µs] | 1e3 | | 100 | 100 | |
| Hit Lockout Time [µs] | 10e3 | | 10e3 | 10e3 | |
| Threshold de-trigger | | 9e-5 | | | |
| STA window time [µs] | | 75 | | | |
| LTA window time [µs] | | 1e6 | | | |
| Pre-event time [µs] | | 15 | | | |
| Post-event time [µs] | | 10e3 | | | |
| Weighting-R constant | | | 4 | 4 | |
| End delay time window 1 [µs] | | | 25 | 25 | |
| End delay time window 2 [µs] | | | 10 | | |
| Start delay time window 1 [µs] | | | | 1.5e3 | |
| Start delay time window 2 [µs] | | | 100 | | |
| CWT scales | | | | 101 | |
| Grayscale image bit-depth | | | | 16 | |
| Median filter pixel neighbors | | | | 50 | |
| STA length [µs] | | | | | 20 |
| STA window | | | | | Hamming |
| Overlapping window samples | | | | | 1 |
| ZCR threshold [%] | | | | | 70 |
| α-weighting STD noise | | | | | 4 |
| Early noise analysis [µs] | | | | | 2e3 |

In **Fig. 11** an instance of the manual onset pick procedure for an AE event is depicted. This is carried out by identifying the time instant when the bi-dimensional manifold created by means of the contour mapping of the SSWT becomes closed by connecting all the modal frequencies of the signal.

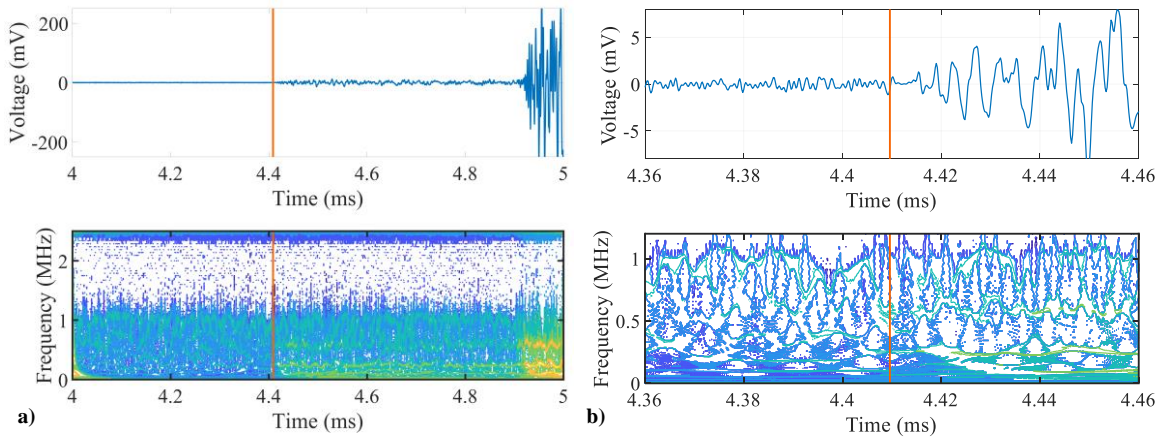

**Fig. 11.** Time-voltage and Time-Frequency representations corresponding to the onset of the synthetic AE wave showed in Fig. 6. (**a**) Data frame of 1ms displaying the onset location (at 4.4084ms, vertical orange line) of the AE wave, and showing the energetic activity of the modal frequencies after the signal arrival in the TFR. (**b**) Zoom of 100µs of the data frame depicting the appearance in the TFR of the most energetic continuous ridge (located at 263.12kHz), that indicates the onset of the AE wave.

In **Fig. 12** and **Fig. 13**, the resulting onset and endpoint automatic procedures for each method are depicted, using in all cases the same AE signal (and which corresponds to **Fig. 6**).

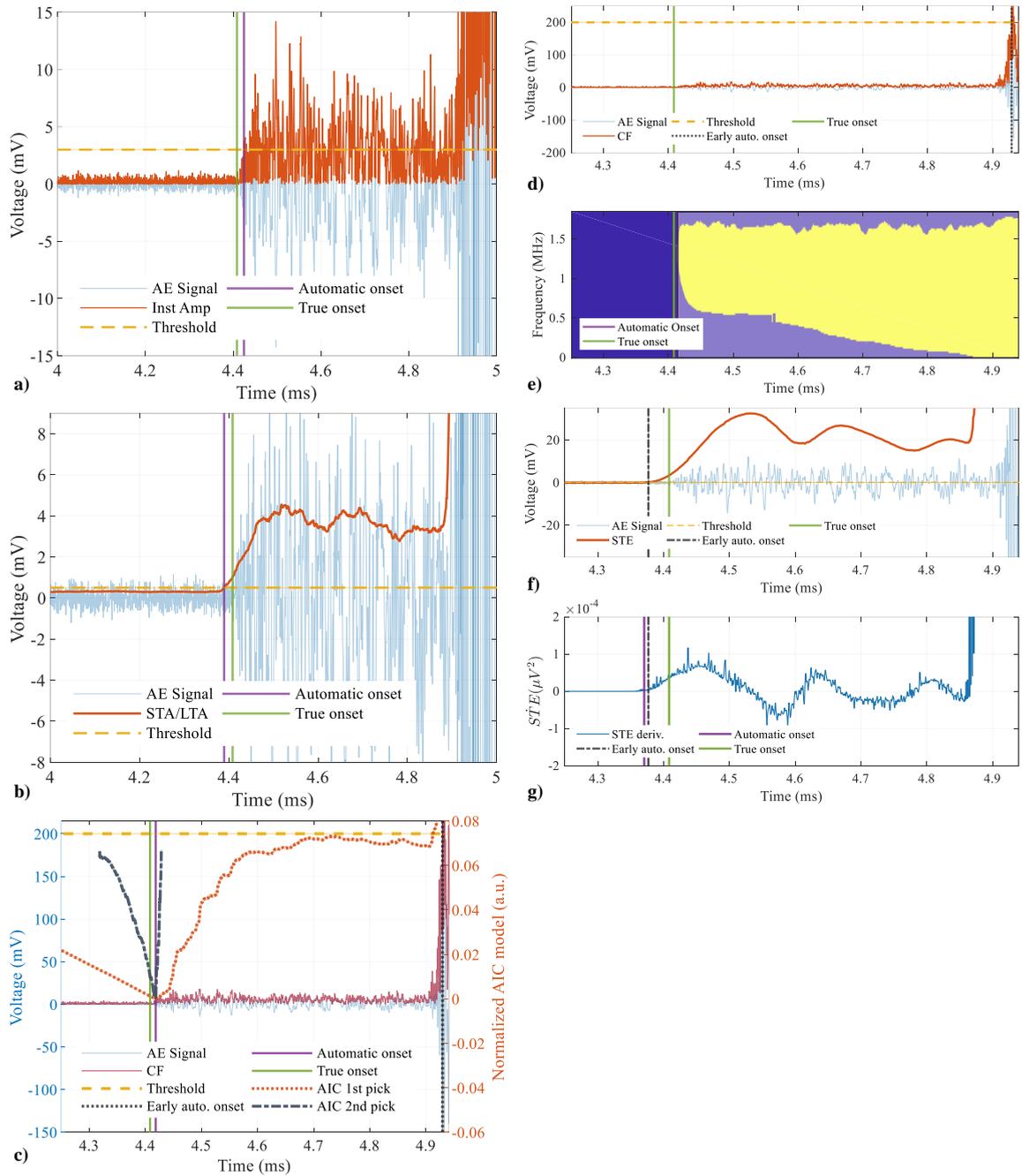

**Fig. 12.** Automatic onset detection procedure of the AE wave, corresponding to the methods: **(a)** Instantaneous amplitude, **(b)** STA/STL, **(c)** two-step AIC, **(d)** CWT-Otsu (early detection), **(e)** CWT-Otsu (detection refinement), **(f)** Short-Time Energy (early detection), **(g)** Short-Time Energy derivative (detection refinement). The onset absolute error corresponding to each method is calculated with regard to the manual True onset pick of the AE wave (vertical green solid line located at 4.4084ms for this instance), and the Automatic onset pick (vertical lilac solid line) that each method identifies.

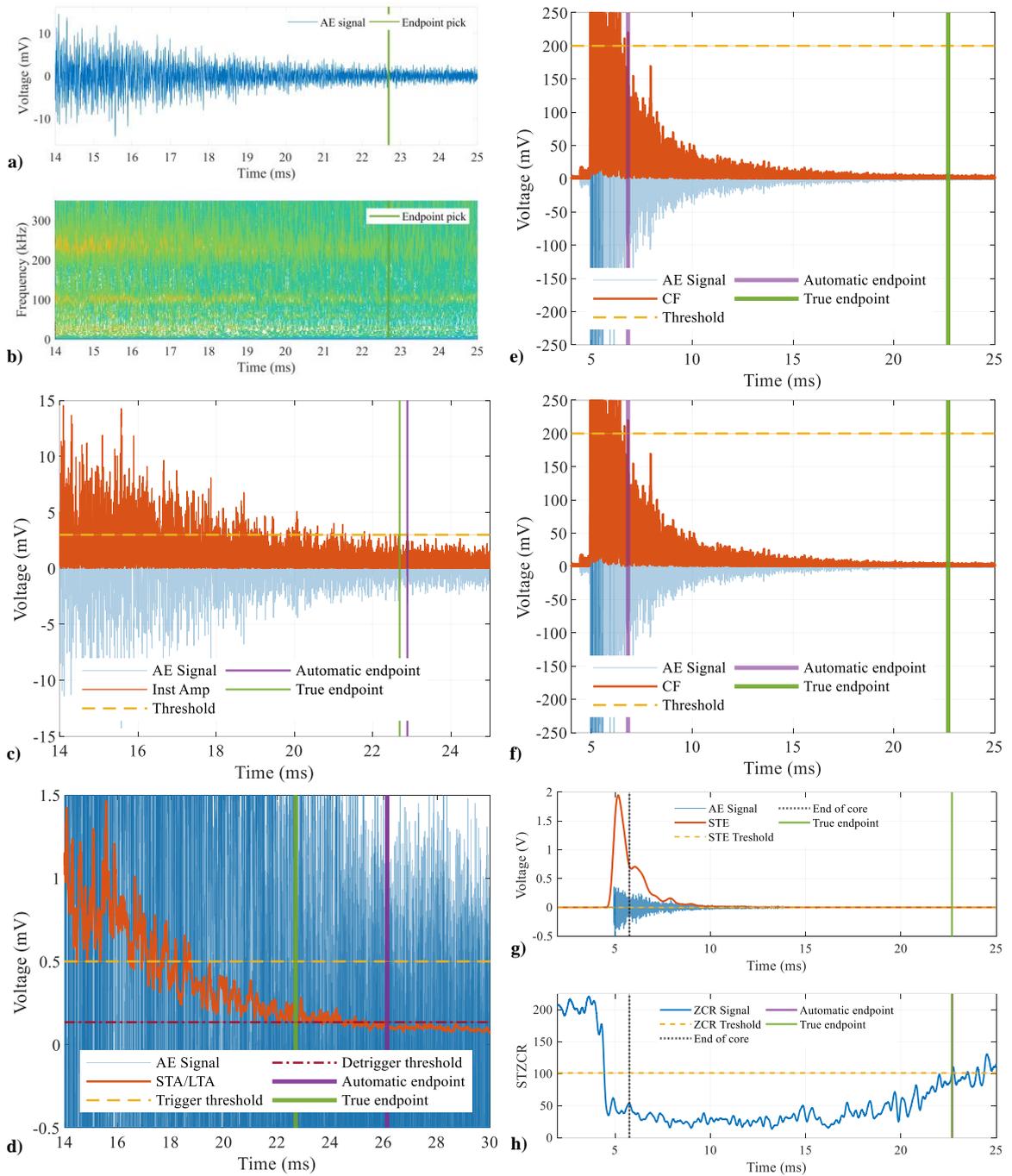

**Fig. 13.** Manual endpoint pick (vertical green solid line located at 22.6973ms for this instance), and detected when the bi-dimensional manifold created by means of the most energetic ridges of the SSWT is vanished, **(a)** Time-Voltage, **(b)** Time-Frequency. Automatic endpoint detection procedures of the AE wave corresponding to the methods: **(c)** Instantaneous amplitude, **(d)** STA/STL, **(e)** two-step AIC, **(f)** CWT-Otsu, **(g)** Short-Time Energy (early endpoint detection) and **(g)** Zero-Crossing Rate (detection refinement). The endpoint absolute error corresponding to each method is calculated with regard to the true endpoint of the AE wave.

For the onset determination, in **Fig. 12** can be observed that due to the significant difference between the amplitudes of the primary wave (4.4 - 4.9ms) and the secondary wave (from 4.9ms on) regarding to the AE artificial source, all methods deal with a challenging signal to accurately determine its onset time. This condition forces to lower down the threshold level as minimum as possible for the IA and the STA/LTA methods (increasing the chances to false-positive detections due to noise floor). For the AIC and CWT-Otsu methods, since they perform an onset refinement detection procedure, they allow to maintain a higher threshold level for the early threshold detection (with the aim to avoid false-positive detections). In the same way, the proposed method through its onset refinement measure by means of the derivative of the STE, also allows to maintaining a higher threshold level in order to avoid trigger false-positive detections.

For the endpoint detection case, **Fig. 13 (h)** shows that with the Zero-Crossing-Rate procedure although in order to be operative still makes use of a threshold value parameter, this measure provides a reliable indicator with which determine the conclusion of the signal. In contrast, **Fig. 13** it also shows that the other methods by making use of a combination of a threshold level along with preset fixed timers entail to a lack of accuracy to the estimation of the end of the signal.

In **Table 2** the accuracy of the onset, endpoint and lifespan of the analyzed methods are quantified by means of the absolute error (from the outcomes of the analyzed methods with regard to the manually picked instants of time). In **Table 2,** it is also showed the accuracy results for the operational robustness in front of three induced levels of background noise on the dataset. Finally, the average required consumed time in order to process an AE event belonging to the dataset are also displayed.

**Table 2**
Absolute error and standard deviation of the onset, endpoint and lifespan detections in regard with the Hsu-Nielsen test-bench.

| Method | Onset error (µS) | Endpoint error (µs) | Lifespan error (µs) | Processing time (s) |
|---|---|---|---|---|
| *IA (original signal)* | **-21.83** ± 8.26 | **2454** ± 1120 | 2476 ± 1120 | **2.53** ± 0.34 |
| *IA (SNR 20dB)* | 4271 ± 110.3 | -20683 ± 1211 | -24955 ± 1197 | 19.30 ± 2.32 |
| *IA (SNR 15dB)* | 4271 ± 110.4 | -20504 ± 1273 | -24776 ± 1263 | 22.94 ± 1.03 |
| *IA (SNR 10dB)* | 4271 ± 110.4 | -20104 ± 1275 | -24376 ± 1268 | 25.98 ± 1.59 |
| *STA/LTA (original signal)* | **-19.82** ± 7.92 | **3828** ± 1159 | 3848 ± 1161 | **1.56** ± 0.12 |
| *STA/LTA (SNR 20dB)* | 27.22 ± 185.44 | 3875 ± 1071 | 3848 ± 1161 | 1.58 ± 0.34 |
| *STA/LTA (SNR 15dB)* | 4256 ± 110.88 | 8104 ± 1174 | 3848 ± 1161 | 1.56 ± 0.11 |
| *STA/LTA (SNR 10dB)* | 4272 ± 110.39 | 8131 ± 1195 | 3902 ± 1197 | 1.60 ± 0.09 |
| *AIC (original signal)* | **-13.34** ± 7.00 | **16338** ± 1045 | 16352 ± 1045 | **2.98** ± 0.14 |
| *AIC (SNR 20dB)* | -515.18 ± 7.31 | 18825 ± 1126 | 19340 ± 1126 | 3.09 ± 0.16 |
| *AIC (SNR 15dB)* | -517.02 ± 6.96 | 18661 ± 1100 | 19178 ± 1100 | 3.24 ± 0.21 |
| *AIC (SNR 10dB)* | 2121 ± 2066 | 21959 ± 3108 | 19838 ± 1152 | 6.60 ± 9.92 |
| *CWT-Otsu (original signal)* | **-1.19** ± 97.88 | **17795** ± 1047 | 17796 ± 1039 | **1.57** ± 0.12 |
| *CWT-Otsu (SNR 20dB)* | 369.29 ± 582.2 | 18881 ± 1119 | 18447 ± 1301 | 1.67 ± 0.14 |
| *CWT-Otsu (SNR 15dB)* | 320.00 ± 647.9 | 18656 ± 1092 | 18336 ± 1309 | 1.85 ± 0.17 |
| *CWT-Otsu (SNR 10dB)* | 2425 ± 1892 | 21632 ± 3274 | 19207 ± 1763 | 4.52 ± 1.55 |
| *STE-ZCR (original signal)* | **3.29** ± 13.40 | **-0.63** ± 1797 | 3.31 ± 1800 | **0.013** ± 0.001 |
| *STE-ZCR (SNR 20dB)* | -29.36 ± 12.13 | 9560 ± 2142 | 9589 ± 2145 | 0.029 ± 0.009 |
| *STE-ZCR (SNR 15dB)* | -22.90 ± 193.7 | 11510 ± 4459 | 11533 ± 4467 | 0.027 ± 0.011 |
| *STE-ZCR (SNR 10dB)* | 236.5 ± 873.66 | 12986 ± 5788 | 12749 ± 5263 | 0.024 ± 0.009 |

For the onset detection measure, the results indicate that all methods perform relatively well, accomplishing errors less than 20µs. However, the CWT-Otsu is the method that achieved the higher accuracy, obtaining an average error of only 1.19µs, yet at expense of displaying the worst dispersion of the considered methods For the STE-ZCR case, it reached the second best results, and also can be observed that the STE-ZCR technique tend to detect the AE event before of its arrival, contrasting with the rest of the methods which are likely to determine the onset time after of the actual start of the event.

For the endpoint detection case, results reveal that none of the comparative methods achieves a reliable measurement, yielding to absolute errors about two and four orders of magnitude with regard to the onset detection procedure. As observed in **Fig. 13**, this lack of accuracy owes that all methods make use of the combination of threshold levels along with preset fixed timers (i.e., HDT, HLT, etc.), without considering the actual behavior of the signal. As consequence, these inaccuracies for the endpoint detection directly lead to errors to the lifespan determination. Differently, and despite of producing the result with higher dispersion and still depending on a calibration parameter, the STE-ZCR method achieved the best accuracy by considering an intrinsic indicator of the waveform.

For the results regarding to the average consumed time to process each AE event, it can be observed that the most expensive technique corresponds to the two-step AIC, since it involves the refinement of the onset measure in two sequential instances (having to modelling the signal in two occasions as consequence). For the IA technique, despite of performing the most basic approach of the considered methods, it achieved the second poorest performance on the test-bench, owing to that it is computationally expensive the searching-and-resetting scheme required to determine the endpoint time over a CF, which exhibits a pronounced amount of rippling. The most balanced options are represented by the STA/LTA and the CWT-Otsu methods, by reducing about to half of the required processing time with regard to AIC and IA techniques. Still, for the STE-ZCR method case, since its operation is carried out on a very straightforward fashion, the results show that its performance by means of the considered software implementation greatly excels to the current methods of the state of the art, achieving a reduction about of 99% of time regarding to the comparative methods.

Finally, the results corresponding to the operational robustness in front of induced AWG background noise, show that the less resilient method is the IA technique by saturating both measurements (i.e., by leading the onset detection to the start of the data frame, and lagging the endpoint determination to the end of said frame), at the first evaluation of added noise; additionally, it also can be noticed that the required processing time considerably increases when a greater amount of rippling is present in the CF. For the STA/LTA method by being endowed with a secondary threshold (aimed to determine the end of the event), their endpoint measurements showed some regularity during for all noise evaluations, additionally, the method proved to be the most resilient for the first evaluation of added noise (due to the robustness delivered by its CF), furthermore, the required processing time was sustained for all evaluations; however, the onset measurement failed for the 15 and 10dB ratios, by showing saturation. The performance achieved by AIC technique for the onset determination revealed some regularity for the 20 and 15dB ratios, nevertheless the measurement failed for the 10dB ratio evaluation; for the endpoint measure, by only depending on the threshold-timer scheme and by maintaining a high threshold level value for the test-bench, the obtained results were closed to those generated when evaluating the original signal, still, all of these containing a high uncertainty degree; it also can be observed that as the signal contained greater presence of AWG noise, the task of modelling the AE signal becomes more difficult for the technique, impacting on the required processing time as consequence. The onset evaluation achieved by the AIC-Otsu method, was the one that showed the closest coherence with regard to the AE phenomenon development, by approaching to correctly determine the arrival of the secondary waves for the 20 and 15dB ratios, nevertheless its accuracy was lost for the 10dB ratio; for the endpoint measure, by implementing the same scheme of the AIC in order to determine the conclusion of an AE event, results are equivalent; for the analysis of the average consumed time per AE event, the technique showed a tolerable time consumption in order to analyze the three levels of induced AWGN with regard to the results obtained by analyzing the original signal. For the case of the STE-ZCR method, owing to implement a threshold adjustment procedure based on an early estimation of noise floor, the technique proved to be the most resilient alternative for onset detection measure regarding to the considered methods, by considerably reducing the amount of error as well as never saturating its detections; for the endpoint determination and despite of been heavily biased by AWGN, the ZCR alternative measure showed coherence with regard to the AE phenomenon development; for the required average processing time, the method proved to be the most efficient alternative by practically maintain unaltered the results obtained by analyzing the original signal.

*4.2 Uniaxial tensile test. Quality of detection statistical indicators.*

The objective for this second test-bench is to quantify the quality for event detection over a data frame collected from a standardized tensile test, which contains a substantial diversity of continuous AE events. In comparison with the artificial AE events produced by the Hsu-Nielsen procedure, real AE waves typically will exhibit smaller amplitudes and shorter durations (of course depending on the stage of damage of the specimen under evaluation). Therefore, for the calibration used for this test-bench (see **Table 3**), the time-driven parameters as well as the threshold levels have been reduced in order to increase the sensitivity (regarding to temporal and amplitude detection capabilities) of the methods.

**Table 3**
Calibration parameters values used for each method for the field data test-bench.

| Parameter | IA | STA LTA | AIC | CWT Otsu | STE ZCR |
|---|---|---|---|---|---|
| Fixed threshold level | 2.25e-3 | 4e-3 | 6e-3 | 6e-3 | 55e-6 |
| Hit Definition Time [μs] | 100 | | 100 | 100 | |
| Hit Lockout Time [μs] | 15 | | 15 | 15 | |
| Threshold de-trigger | | 3e-3 | | | |
| STA window time [μs] | | 25 | | | |
| LTA window time [μs] | | 10e3 | | | |
| Pre-event time [μs] | | 1 | | | |
| Post-event time [μs] | | 0.5 | | | |
| Weighting-R constant | | | 4 | 4 | |
| End delay time window 1 [μs] | | | 10 | 10 | |
| End delay time window 2 [μs] | | | 5 | | |
| Start delay time window 1 [μs] | | | | 75 | |
| Start delay time window 2 [μs] | | | 20 | | |
| CWT scales | | | | 101 | |
| Grayscale image bit-depth | | | | 16 | |
| Median filter pixel neighbors | | | | 50 | |
| STA duration [μs] | | | | | 15 |
| STA window | | | | | Hamming |
| Overlapping window samples | | | | | 1 |
| ZCR threshold [%] | | | | | 80 |
| α-weighting STD noise | | | | | 1 |
| Early noise analysis [μs] | | | | | 5 |

Once that all methods have processed the field data frame of 500ms length, the quality of event detection is quantified in two steps. First, by totaling the total number of detected events against the true locations (referring a total amount of 380 AE events present in the data frame), and by classifying the properly detected events (true-positive), missed events (false-negative) and the mistakenly detected events (false-positive), that each method concludes (see **Table 4**).

**Table 4**
Classification of the identified events regarding to 380 AE waves present in the data frame.

|  | Method | | | | |
|---|---|---|---|---|---|
|  | IA | STA/LTA | AIC | CWT-Otsu | STE-ZCR |
| **Total detections** | 373 | 380 | 372 | 372 | 367 |
| **True-positive** | 322 | 324 | 299 | 299 | 338 |
| **False-negative** | 58 | 56 | 81 | 81 | 42 |
| **False-positive** | 51 | 56 | 73 | 73 | 29 |

For this field data test-bench, and from the total number of True-positive detected events, the absolute error for the onset, endpoint and lifespan are also calculated (see **Table 5**).

**Table 5**
Absolute error and standard deviation of the onset, endpoint and lifespan detections regarding to the field data test-bench.

| Method | Onset error (µs) | Endpoint error (µs) | Lifespan error (µs) |
|---|---|---|---|
| IA | -9.69 ± 7.56 | 38.39 ± 101.27 | 48.09 ± 102.13 |
| STA/LTA | -2.49 ± 8.63 | 12.07 ± 83.65 | 14.56 ± 84.85 |
| AIC | -6.15 ± 10.44 | 19.57 ± 543.74 | 50.4 ± 692.6 |
| CWT-Otsu | 2.53 ± 29.45 | -92.36 ± 97.4 | 89.82 ± 97.74 |
| STE-ZCR | 4.62 ± 53.79 | -4.42 ± 114.15 | -9.05 ± 102.04 |

From **Table 5**, it can be observed that despite of reducing the onset and endpoint error measures with regard to the Hsu-Nielsen test-bench (due to having to deal with less challenging AE events by exhibiting less pronounced s-waves), the order regarding to the competency of the performance of the considered methods is better aligned with the results obtained for the first round of operative robustness (SNR 20dB). All methods exhibit a sustained performance for the onset detection of the AE events by achieving an average error less than 10µs for all cases. For this experimental scenario, the STE-ZCR method accomplishes yet again the best endpoint determination and in consequence the best lifespan measure, ensuring that most of the detected events are complete in case of requiring a subsequent assessing analysis. From results of **Table 4**, it can be observed that despite that in average, all the comparative methods nearly detect 99% of the detection objective (i.e., 380-hits), none of them reaches more than 85% of true-positive detections. In consequence, given the quality with which these detections are performed, the reliability of the methods is not assured. In **Fig. 14**, a representative instance (which is derived of the test-bench and containing five different AE events) depicting this matter, as well as the different automatic detections that each method identifies, is showed.

As it can be observed, none of the analyzed methods performs the required detections without errors. Such is the case of the Instantaneous Amplitude method, which after the first event detection it splices four AE events as if it was a single one, this error owes to the use of the threshold level along with the fixed HDT for the endpoint detection corresponding to the second event. For the STA/LTA method, its CF helps to depict better the dynamic of the phenomenon and therefore to detecting more events, nevertheless, by implementing the same procedure of a threshold level and a fixed timer, it splices the second and third event as a single detection; finally, the method also executes a false-positive detection corresponding to a wave reflection of the last event. In this instance, it is clear that AIC and CWT-Otsu methods necessarily require a precise early onset estimation in order to achieve a reliable onset automatic refinement, since despite of accurately detect the first event, due to the configuration of the window length for the early onset the refinement procedure the less energetic events are discarded (i.e., the second, third and fourth events), by detecting the fifth AE event as the start of the wave (and therefore the only existing event of the data-frame), and leading to three false-negative detections as consequence.

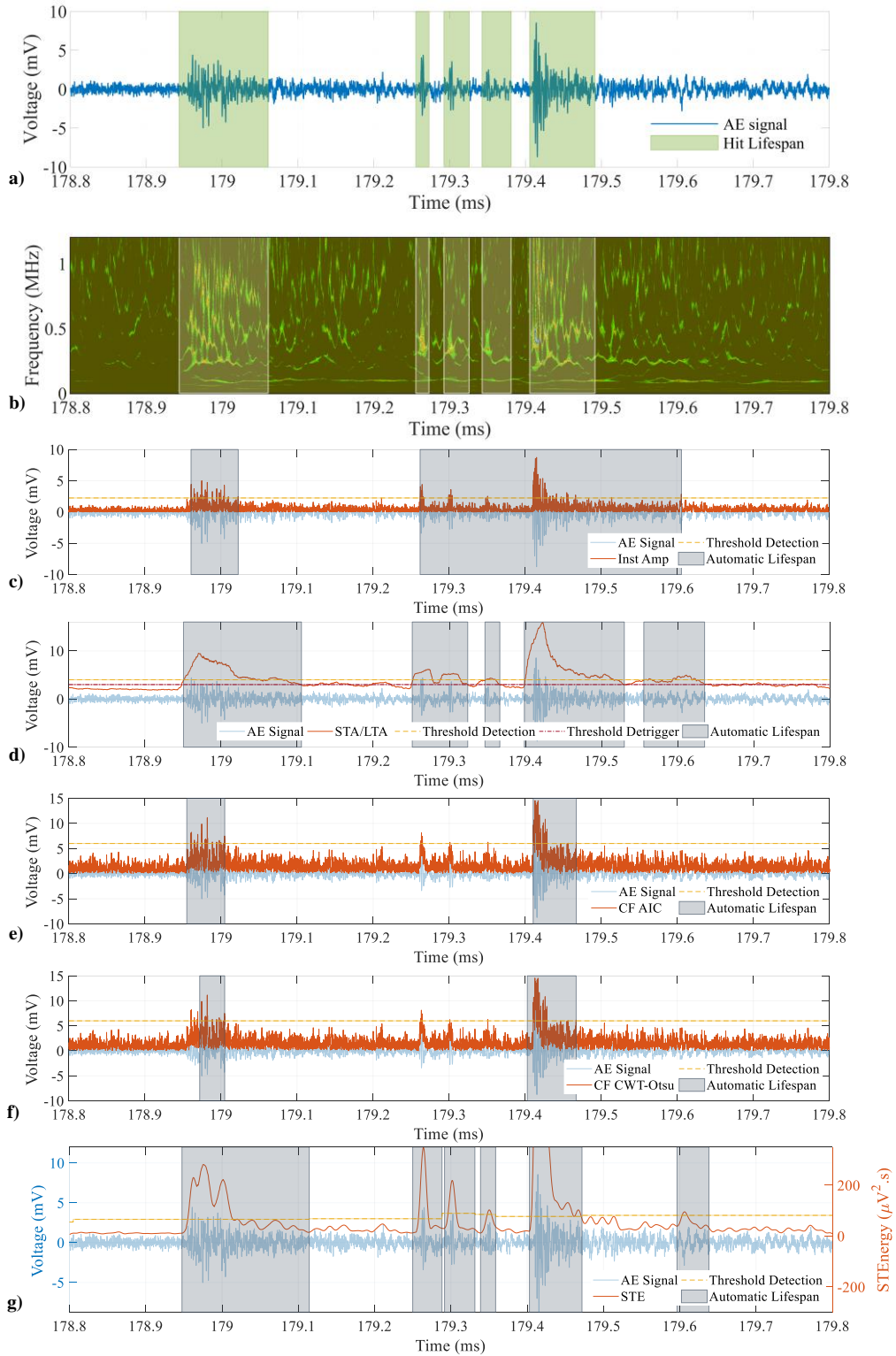

**Fig. 14.** Comparison of events detected by the considered methods for an instance of 1ms (derived from the uniaxial tensile test). **(a)** Time-Voltage and **(b)** Time-Frequency manual detections. Automatic: **(c)** Instantaneous Amplitude, **(d)** STA/LTA, **(e)** AIC, **(f)** CWT-Otsu and **(g)** STE-ZCR proposed technique.

Finally, the STE-ZCR method despite of also executing the same false-positive detection as the STA/LTA technique (corresponding to the last event at 179.6ms), it can be observed that is the only method that accomplishes the detection of all existing events, besides of also achieving the most accurate lifespan measure for said events.

The last step for this test-bench comprises on the quantification of the quality of event detection of the considered methods. For this, by means of the amount of classified events from **Table 4**, a set of statistical indicators are calculated. These are: (a) accuracy (the ratio of true-positive events to all detected and not detected events), (b) precision (the ratio of true-positive events to the amount of true and false-positive events), (c) sensitivity (the ratio of true positive events to the sum of true-positive and false-negative detections), (d) F1-score (the harmonic average between precision and sensitivity), (e) false discovery rate (the ratio of false-positive detections to all detected events), (f) false-negative rate (the ratio of false-negative detections to the sum of false-negative and true-positive events), see **Table 6**.

**Table 6**
Statistical metrics corresponding to the quality for event detection concerning to the data field test-bench.

|  | Method | | | | |
| --- | --- | --- | --- | --- | --- |
|  | *IA* | *STA/LTA* | *AIC* | *CWT Otsu* | *STE ZCR* |
| **Accuracy (%)** | 74.71 | 74.31 | 66.00 | 66.00 | 82.64 |
| **Precision (%)** | 86.33 | 85.26 | 80.38 | 80.38 | 92.10 |
| **Sensitivity (%)** | 84.74 | 85.26 | 78.68 | 78.68 | 88.95 |
| **F1 score (%)** | 85.52 | 85.26 | 79.52 | 79.52 | 90.50 |
| **False discovery rate (%)** | 13.67 | 14.74 | 19.62 | 19.62 | 7.90 |
| **False negative rate (%)** | 15.26 | 14.74 | 21.32 | 21.32 | 11.05 |
| **Processing time (sec.)** | 220.49 | 33.54 | 608.56 | 182.92 | 19.6 |

For this test-bench, it can be observed that comparative methods perform with reasonable accuracy confidence by achieving an average score of 70%, and also can be noticed that the accuracy score is consistently aligned with the endpoint determination accuracy (i.e., the better the endpoint is detected, the higher the accuracy will score). In general terms all methods achieve better scores for the precision indicator than for the accuracy, this owes to the fact that in the dataset exists a larger amount of true AE events with regard to false events (as high-energy reflections and mechanical noises, e.g., slips). For the sensitivity metric, all methods slightly diminish their performances with regard to precision metric by being slightly prone to generate false-negative detections (mostly of them derived from spliced detections).

In the case of the STE-ZCR method, as can be observed it excels to the rest of the considered techniques about 12% for the accuracy metric, 10% for the precision metric, 7% for the sensitivity metric, 8% for the F1-score, 10% for the false discovery rate and 7% for the false negative rate case. This improvement clearly owes to performing the endpoint detection with higher accuracy. Finally, it can also be observed that the STE-ZCR technique reduces the required processing time with regard of the rest of the considered methods about 45-97%.

## 5. Conclusions

An Acoustic Emission activity detector, which allows an automatic and continuous detection of AE events, was developed using time domain features obtained from the waveform of the signal of interest. The proposed methodology was realized by revisiting a well-established signal processing technique from the speech processing area, and adapting it to the requirements of the AE phenomenon.

Two experimental scenarios were arranged to quantifying the performance of the proposed method, and with the aim to analyzing three critical aspects related to the AE event detection: the onset and endpoint accuracies, as well the quality of event detection.

Firstly, in the Hsu-Nielsen test-bench, for the case of the onset detection measure, the proposed STE-ZCR method improved the accuracy with regard to the IA method by in average, diminishing the measuring error by 18.54µs, 16.53µs for the STA/LTA and 10.05µs for the AIC method. The STE-ZCR method despite of producing a larger average error of 2.1µs with respect to the CWT-Otsu technique, it diminished the respective dispersion error by 84.48µs.

For the endpoint detection measure, by implementing an intrinsic indicator derived from the waveform of the AE signal, the STE-ZCR method surpassed the accuracy of the comparative methodologies in about four and five orders of magnitude, which in turn contributed to achieving the lowest error for the lifespan measure.

Lastly for the Hsu-Nielsen test, the STE-ZCR by implementing an adaptive threshold scheme, proved to be the most resilient method under noisy scenarios by never saturating its measurements in the corresponding test.

Secondly, for the uniaxial test-bench, besides of having verified that the obtained results for the onset and endpoint measures of the considered methods are consistent with the obtained for the Hsu-Nielsen test, it was studied the quality with that the AE events are detected.

Results showed that for a data-frame which contains 380 AE events of different durations, amplitudes and random manifestation, the STE-ZCR was the method that achieved the highest amount of true-positive detections and the lowest amount of false (both, positive and negative) identifications. Quantitatively, the STE-ZCR method excelled in average to the rest of the considered techniques in about 12% for the detection accuracy, 10% for the precision and 7% for the sensitivity; and simultaneously reducing in average around of 10% and 7% the false discovery and the false negative rate cases respectively. Once again, this improvement over the rest of the considered methods responds to the fact of implementing a dedicated indicator with which to detect the end of the AE events.

In addition, the proposed STE-ZCR method by using a direct processing scheme, in both experimental scenarios it was the one that achieved the best processing times, by in average reducing the required search time in one up to three orders of magnitude in comparison with the others considered techniques.

It must be noted that despite of the detector has been oriented specifically on the automated detection of AE events for an application related to the characterization of metallic components, by only requiring the waveform of the AE phenomenon to be operative, its direct use for another applications and materials with similar characteristic for their waveforms could be feasible.

Finally, its straightforward scheme and the diminished consumption times, suggest a possible and efficient hardware implementation for online monitoring applications. Furthermore, and in the same line, with the aim of reducing the payload required to transmit or store the large data streams demanded by the phenomenon, by achieving an adequate identification and separation of the AE events, it could be possible to only working with the identified events.

**Acknowledgments**

This work was partially supported by the CONACyT (Consejo Nacional de Ciencia y Tecnología, México) with the scholarship 411711; and by the Spanish Ministry of Economy and Competitiveness, under the TRA2016-80472-R Research Project.